\documentstyle[11pt,epsfig]{article}
\title{\bf Bulk scalar field in brane-worlds with induced gravity inspired by the ${\cal L}(R)$ term}
\author{M. Heydari-Fard$^{1}$ \thanks{email: m.heydarifard@mail.sbu.ac.ir}  \thanks{email: heydarifard@qom.ac.ir}
and H. R. Sepangi$^{2}$ \thanks{email: hr-sepangi@sbu.ac.ir}
\\ {\small Department
of Physics, Shahid Beheshti University, Evin, Tehran 19839,
Iran}$^{2}$\\
{\small Department of Physics, The University of Qom, Qom
37185-359, Iran}$^{1}$}
\begin{document}
\maketitle 
\begin{abstract}
We obtain the effective field equations in a brane-world scenario
within the framework of a DGP model where the action on the brane is
an arbitrary function of the Ricci scalar, ${\cal L}(R)$, and the
bulk action includes a scalar field in the matter Lagrangian. We
obtain the Friedmann equations and acceleration conditions in the
presence of the bulk scalar field for the $R^n$ term in
four-dimensional gravity.
\vspace{5mm}
\end{abstract}
\section{Introduction}
The type Ia supernovae (SNe Ia) \cite{0} observations provide the
first evidence for the accelerating expansion of the present
universe. These results, when combined with the observations on
the anisotropy spectrum of cosmic microwave background (CMB)
\cite{1} and the results on the power spectrum of the large scale
structure (LSS) \cite{2}, strongly suggest that the universe is
spatially flat and dominated by a component, though arguably
exotic, with large negative pressure, referred to as dark energy
\cite{3}. The nature of such dark energy constitutes an open and
tantalizing question connecting cosmology and particle physics.
Different mechanisms have been suggested over the past few years
to accommodate dark energy. The simplest form of dark energy is
the cosmological constant. However, it suffers from serious
problems such as the fine-tuning problem and the coincidence
problem.

An interesting way of explaining the observed acceleration of the
late time universe is to modify gravity at large scales. This
scenario was proposed by Dvali, Gabadadze and Porrati (DGP)
\cite{Dvali}. The DGP proposal is based on the key assumption of the
presence of a four-dimensional $(4D)$ Ricci scalar in the bulk
action. There are two main reasons that make this model
phenomenologically appealing. First, it predicts that $4D$ Newtonian
gravity on a brane-world is regained at distances shorter than a
given crossover scale $r_c$ (high energy limit), whereas $5D$
effects become manifest above that scale (low energy limit)
\cite{Gabadadze}. Second, the model can explain the late time
acceleration without having to invoke a cosmological constant or
quintessential matter \cite{Deffayet,Deffayet1}. For a recent and
comprehensive review of the phenomenology of DGP cosmology, the
reader is referred to \cite{DGP}. Recently, it has been shown that a
very tiny correction to the usual gravitational action of general
relativity of the form $R^n$, with $n<0$ could give rise to
accelerating solutions of the field equations without dark energy
\cite{fR}. In this frame work, some attempts have been made to
explain the observed cosmic acceleration by modifying the
Einstein-Hilbert action.

One important extension to brane-world models is to consider matter
fields in the bulk space, such as scalar fields, see \cite{4} and
references therein. In fact, it is natural that the $5D$ theory is
itself an effective theory which originates from a yet
higher-dimensional theory and the $5D$ effective action includes
some scalar fields of gravitational origin. It is believed that in
the unified theory approach, a dilatonic gravitational scalar field
term is required in the the $5D$ Einstein-Hilbert action \cite{6}.
One of the first motivations for introducing a bulk scalar field has
been to stabilize the distance between the two branes \cite{7} in
the context of the Randall-Sundrum type I brane model (RSI). A
second motivation for studying scalar fields in the bulk is due to
the possibility that such a setup could provide some clue to the
solution of the famous cosmological constant problem. Interestingly,
it has also been shown that inflation is caused solely by the
dynamics of a $5D$ scalar field without introducing it in the brane
universe \cite{sasaki}. The creation of a brane-world with a bulk
scalar field using an instanton as the solution of the $5D$
Euclidean Einstein equations was considered in \cite{9}. Bulk scalar
fields in a DGP brane-world scenario without curvature corrections
have been studied in \cite{India}.

The effective gravitational equations in a brane-world scenario with
induced gravity have been derived in \cite{maeda}. Very recently
Saavedra and Vasquez \cite{joel} have obtained the effective
equations on the brane for modified induced gravity, in this
connection also see \cite{sepangi}. In this paper, we extend the
effective equations derived in \cite{joel} to the case where the
bulk space is endowed with a scalar field. In other words, we study
a bulk scalar field in a DGP brane-world scenario with curvature
correction and show that under certain conditions the universe
undergoes a self-accelerating phase.
\section{Bulk scalar field in DGP model with ${\cal L}(R)$ brane action}
In this section we present a brief review of the model proposed in
\cite{joel} and extend it to the case with a bulk scalar field.
Consider a $5D$ space-time with a $4D$ brane, located at $Y(X^A)=0$,
where $X^A$, ($A=0,1,2,3,4$) are the $5D$ coordinates. The effective
action is given by
\begin{eqnarray}
{\cal S} = \int d^{5}X\sqrt{-{\cal
G}}\left[\frac{1}{2\kappa_5^2}{\cal
R}+S_{m}^{(5)}\right]+\int_{Y=0}d^{4}x
\sqrt{-g}\left[\frac{1}{\kappa_5^2}K^{\pm}+S_{brane}(g_{\alpha\beta},\psi)\right],\label{1}
\end{eqnarray}
where $\kappa_5^2=8\pi G_5$ is the $5D$ gravitational constant,
${\cal R}$ and $S_{m}^{(5)}$ are the $5D$ scalar curvature and the
matter Lagrangian in the bulk, respectively.  Also, $x^{\mu}$
($\mu=0,1,2,3$) are the induced $4D$ coordinates on the brane,
$K^{\pm}$ is the trace of the extrinsic curvature on either side of
the brane \cite{G} and $S_{brane}(g_{\alpha\beta},\psi)$ is the
effective $4D$ Lagrangian, which is given by a generic functional of
the brane metric $g_{\alpha\beta}$ and matter fields.

The $5D$ Einstein field equations are given by
\begin{eqnarray}
{\cal R}_{AB}-\frac{1}{2}{\cal R}{\cal
G}_{AB}={\kappa_5^2}\left[T^{(5)}_{AB}+\delta(Y)\tau_{AB}\right],\label{2}
\end{eqnarray}
where
\begin{eqnarray}
T^{(5)}_{AB}\equiv-2\frac{\delta S^{(5)}_{m}}{\delta{\cal
G}^{AB}}+{\cal G}_{AB}S^{(5)}_{m},\label{3}
\end{eqnarray}
and
\begin{eqnarray}
\tau_{\mu\nu}\equiv-2\frac{\delta S_{brane}}{\delta
g^{\mu\nu}}+g_{\mu\nu}S_{brane}.\label{4}
\end{eqnarray}
We study the case where the induced gravity scenario arises from
higher-order corrections to the scalar curvature over the brane. The
interaction between the bulk gravity and local matter induces
gravity on the brane through its quantum effects. If we take into
account quantum effects of the matter fields confined to the brane,
the gravitational action on the brane is modified as
\begin{eqnarray}
S_{brane}(g_{\alpha\beta},\psi) = \frac{\mu^2}{2}{\cal
L}(R)-\lambda+S_{m},\label{5}
\end{eqnarray}
where $\mu$ is a mass scale which may correspond to the $4D$ Planck
mass, $\lambda$ is the tension of the brane and $S_m$ presents the
Lagrangian of the matter fields on the brane. We note that for
${\cal L}(R)=R$, action (\ref{1}) gives the DGP model if $\lambda=0$
and $\Lambda^{(5)}=0$ and gives the RSII model if $\mu=0$.

We obtain the gravitational field equations on the brane-world as
\cite{Shiromizu}
\begin{eqnarray}
G_{\mu\nu} = \frac{2\kappa_5^2}{3}\left[T_{AB}^{(5)}g^A_{\mu}
g^B_{\nu}+g_{\mu\nu}\left(T_{AB}^{(5)}n^An^B-\frac{1}{4}T^{(5)}\right)\right]+\kappa_{5}^4\pi_{\mu\nu}-{\cal
E}_{\mu\nu},\label{6}
\end{eqnarray}
\begin{eqnarray}
\nabla_{\nu}\tau_{\mu}^{\nu} =
-2T_{AB}^{(5)}n^{A}g^{B}_{\mu},\label{7}
\end{eqnarray}
where $\nabla_{\upsilon}$ is the covariant derivative with respect
to $g_{\mu\nu}$ and the quadratic correction has the form
\begin{eqnarray}
\pi_{\mu\nu} =
-\frac{1}{4}\tau_{\mu\alpha}\tau_{\nu}^{\alpha}+\frac{1}{12}\tau\tau_{\mu\nu}+
\frac{1}{8}g_{\mu\nu}\tau^{\alpha\beta}\tau_{\alpha\beta}-\frac{1}{24}g_{\mu\nu}\tau^2,\label{8}
\end{eqnarray}
and the projection of the bulk Weyl tensor to the surface
orthogonal to $n^A$ is given by
\begin{eqnarray}
{\cal E}_{\mu\nu} = C^{(5)}_{ABCD}n^An^Bg^C_{\,\,\,\mu}
g^D_{\,\,\,\nu}.\label{9}
\end{eqnarray}
In order to find the basic field equations on the brane with induced
gravity described by the ${\cal L}(R)$ term, we have to obtain the
energy-momentum tensor of the brane $\tau_{\mu\nu}$, given by
definition (\ref{4}) from Lagrangian (\ref{5}), yielding
\begin{eqnarray}
\tau_{\mu\nu} =
-\Lambda(R)g_{\mu\nu}+T_{\mu\nu}-\Sigma(R)G_{\mu\nu}+D_{\mu\nu},\label{10}
\end{eqnarray}
where the functions $\Lambda(R)$, $\Sigma(R)$ and $D_{\mu\nu}$ are
defined as
\begin{equation}\label{11}
\Lambda(R) = \frac{\mu^2}{2}\left[R\frac{d{\cal L}(R)}{dR}-{\cal
L}(R)+2\frac{\lambda}{\mu^2}\right],
\end{equation}
and
\begin{equation}\label{12}
\Sigma(R) = \mu^2\frac{d{\cal L}(R)}{dR},
\end{equation}
and
\begin{equation}\label{13}
D_{\mu\nu} = \mu^2\left[\nabla_\mu \nabla_\nu{\left(\frac{d{\cal
L}(R)}{dR}\right)}-g_{\mu\nu}\nabla^\beta
\nabla_\beta\left(\frac{d{\cal L}(R)}{dR}\right)\right].
\end{equation}
Let us now introduce a scalar field in the bulk and assume that the
cosmological constant is zero $\Lambda^{(5)}=0$. The energy momentum
tensor of the bulk scalar field is given by
\begin{equation}\label{14}
T^{(5)}_{AB} = \phi_{,A}\phi_{,B}-{\cal
G}_{AB}\left(\frac{1}{2}{\cal
G}^{CD}\phi_{,C}\phi_{,D}+V(\phi)\right).
\end{equation}
Inserting equations (\ref{10}) and (\ref{14}) into equation
(\ref{6}), we find the effective field equations for the $4D$ metric
$g_{\mu\nu}$ as
\begin{eqnarray}
[1&+&\frac{1}{6}\kappa_{5}^{4}\Lambda(R)\Sigma(R)]G_{\mu\nu} =
\frac{1}{6}\kappa_{5}^{4}\Lambda(R)T_{\mu\nu}+\kappa_{5}^2\hat{T}_{\mu\nu}-\frac{1}{12}\kappa_{5}^{2}\Lambda(R)^2
g_{\mu\nu}+
\frac{1}{6}\kappa_{5}^{4}D_{\mu\nu}+\kappa_{5}^{4}\pi_{\mu\nu}^{(T)}+\mu^4\pi_{\mu\nu}^{(D)}\nonumber\\
\nonumber\\
&+&\kappa_{5}^{4}\Sigma(R)^2\pi_{\mu\nu}^{(G)}-\kappa_{5}^{4}\Sigma(R){\cal
K}^{(T)}_{\mu\nu\alpha\beta}G^{\alpha\beta}-\kappa_{5}^{4}\left(\Sigma(R)G^{\alpha\beta}-T^{\alpha\beta}\right){\cal
K}^{(D)}_{\mu\nu\alpha\beta}-{\cal E}_{\mu\nu},\label{15}
\end{eqnarray}
where
\begin{equation}\label{16}
\hat{T}_{\mu\nu} =
\frac{1}{6}\left[4\phi_{,\mu}\phi_{,\nu}+\left(\frac{3}{2}(\phi_{,y})^2-
\frac{5}{2}g^{\alpha\beta}\phi_{,\alpha}\phi_{,\beta}-3V(\phi)\right)
g_{\mu\nu}\right],
\end{equation}
\begin{eqnarray}
\pi^{(T)}_{\mu\nu} =
-\frac{1}{4}T_{\mu\alpha}T^{\alpha}_{\nu}+\frac{1}{12}TT_{\mu\nu}
+\frac{1}{8}g_{\mu\nu}T_{\alpha\beta}T^{\alpha\beta}-\frac{1}{24}g_{\mu\nu}T^2,\label{17}
\end{eqnarray}
\begin{eqnarray}
\pi^{(G)}_{\mu\nu} =
-\frac{1}{4}G_{\mu\alpha}G^{\alpha}_{\nu}+\frac{1}{12}GG_{\mu\nu}
+\frac{1}{8}g_{\mu\nu}G_{\alpha\beta}G^{\alpha\beta}-\frac{1}{24}g_{\mu\nu}G^2,\label{18}
\end{eqnarray}
\begin{eqnarray}
\pi^{(D)}_{\mu\nu} =
-\frac{1}{4}D_{\mu\alpha}D^{\alpha}_{\nu}+\frac{1}{12}DD_{\mu\nu}
+\frac{1}{8}g_{\mu\nu}D_{\alpha\beta}D^{\alpha\beta}-\frac{1}{24}g_{\mu\nu}D^2,\label{19}
\end{eqnarray}
and
\begin{eqnarray}
{\cal K}^{(T)}_{\mu\nu\rho\sigma} =
\frac{1}{4}\left(g_{\mu\nu}T_{\rho\sigma}-g_{\mu\rho}T_{\upsilon\sigma}-g_{\nu\sigma}T_{\mu\rho}\right)+\frac{1}{12}
\left[T_{\mu\nu}g_{\rho\sigma}+T(g_{\mu\rho}g_{\nu\sigma}-g_{\mu\nu}g_{\rho\sigma})\right],\label{20}
\end{eqnarray}
\begin{eqnarray}
{\cal K}^{(D)}_{\mu\nu\rho\sigma} =
\frac{1}{4}\left(g_{\mu\nu}D_{\rho\sigma}-g_{\mu\rho}D_{\upsilon\sigma}-g_{\nu\sigma}D_{\mu\rho}\right)+\frac{1}{12}
\left[D_{\mu\nu}g_{\rho\sigma}+D(g_{\mu\rho}g_{\nu\sigma}-g_{\mu\nu}g_{\rho\sigma})\right],\label{21}
\end{eqnarray}
with $T$ being the trace of the energy momentum tensor and $D$ is
\begin{equation}\label{22}
D = g^{\mu\nu}D_{\mu\nu} =
-3\mu^2\nabla^{\alpha}\nabla_{\alpha}\left(\frac{d{\cal
L}(R)}{dR}\right).
\end{equation}
We note that these equations are exactly the same effective
equations as presented in reference \cite{joel} in the absence of
a bulk scalar field. In the next section, we discuss the effects
of a bulk scalar field on the cosmology of our model.

\section{Cosmology of the model}
A brane-world model with induced gravity described by a ${\cal
L}(R)$ term has a self-accelerated branch in the low energy limit
\cite{joel}. In the following we obtain the effective Friedmann
equation for a Minkowski bulk $\Lambda^{(5)}=0$, ${\cal E}_{00}=0$,
and investigate whether it is possible to have a late time
accelerating phase on the brane when there is a scalar field in the
bulk and the brane is empty.

Using equation (\ref{15}), in the absence of the bulk matter
field, the effective Friedmann equation for a
Friedmann-Robertson-Walker (FRW) universe with zero tension is
then given by \cite{joel}
\begin{eqnarray}\label{joel}
H^2+\frac{k}{a^2} = \frac{1}{4r_c^2{\cal L}'(R)^2}\left[
1+\varepsilon\sqrt{1+\frac{4}{3\mu^2}r_c^2{\cal
L}'(R)\rho_{m}+\frac{2}{3}r_c^2{\cal L}'(R)\left(R {\cal
L}'(R)-{\cal L}(R)-6H\dot{R}{\cal L}''(R)\right)}\right]^2,
\end{eqnarray}
where ${\cal L}'(R)\equiv\frac{d{\cal L}(R)}{dR}$,
$2r_c\equiv\frac{\kappa_5^2}{\kappa_4^2}=\kappa_5^2\mu^2$ and
$\rho_m$ is the energy density of the ordinary matter on the brane
which has a perfect fluid form. The two different possible
$\varepsilon$ namely $\varepsilon=\pm1$, correspond to two different
embedding of the brane into the bulk spac-time. We define $\hat{r}_c
= r_c{\cal L}'(R)$ and $\rho_{tot}=\hat{\rho}_m+\rho_{curv}$, thus
equation (\ref{joel}) can be rewritten as
\begin{equation}\label{n1}
H^2+\frac{k}{a^2} = \left[
\frac{1}{2\hat{r}_c}+\frac{\varepsilon}{2\hat{r}_c}\sqrt{1+\frac{4}{3\mu^2}\hat{r}_c^2\rho_{tot}}\right]^2,
\end{equation}
where
\begin{equation}\label{n2}
\rho_{curv} = \frac{\mu^2}{2{\cal L}'(R)}\left(R {\cal
L}'(R)-{\cal L}(R)-6H\dot{R}{\cal L}''(R)\right),
\end{equation}
and
\begin{equation}\label{n3}
\hat{\rho}_m = \frac{\rho_m}{{\cal L}'(R)}.
\end{equation}
Equation (\ref{n1}) has a similar form to the Friedmann equation in
the DGP model \cite{Deffayet}, differing only in replacing
$\hat{r}_c$ by $r_c$. Since ${\cal L}(R)$ is an arbitrary function
of the Ricci scalar on the DGP brane, different choices of ${\cal
L}(R)$ lead to different forms of $\hat{r}_c$. Also for a specific
function of ${\cal L}(R)$ when it varies from point to point on the
DGP brane, the crossover scale takes different values. By using the
most recent Supernovae observations, the best-fit value for the
crossover scale $r_c$ in terms of the Hubble radius is given by
\cite{alcaniz}
\begin{equation}\label{n3}
r_c \simeq 1.09H_0^{-1}.
\end{equation}
Since ${\cal L}'(R)=\frac{2\kappa_4^2}{\kappa_5^2}\hat{r}_c$, if we
choose $\hat{r}_c \sim 1.09H_0^{-1}$ it allows us to put constraints
on such a ${\cal L}(R)$ scenario, in agreement with observational
data. Using equation (\ref{15}) the effective Friedmann equation
with a bulk scalar field is given by
\begin{eqnarray}\label{23}
H^2+\frac{k}{a^2} = \frac{1}{2r_c^2{\cal
L}'(R)^2}\left\{1+\frac{2}{3\mu^2}r_c^2{\cal
L}'(R)\rho_{m}+\frac{1}{3}r_c^2{\cal L}'(R) \left(R {\cal
L}'(R)-{\cal
L}(R)-6H\dot{R}{\cal L}''(R)\right)\right.\nonumber\\
+\left. \varepsilon\sqrt{1-\frac{4}{3}\kappa_5^2r_c^2{\cal L}'(R)^2
\rho_{\phi}+\frac{4}{3\mu^2}r_c^2{\cal
L}'(R)\rho_{m}+\frac{2}{3}r_c^2{\cal L}'(R)\left(R {\cal
L}'(R)-{\cal L}(R)-6H\dot{R}{\cal L}''(R)\right)}\right\}.
\end{eqnarray}
The energy density and pressure of the bulk scalar field on the
brane are given by
\begin{equation}\label{24}
\rho_{\phi} =
\frac{1}{2}\left[\frac{1}{2}\dot{\phi}^2+V(\phi)\right],
\end{equation}
\begin{equation}\label{25}
P_{\phi} =
\frac{1}{2}\left[\frac{5}{6}\dot{\phi}^2-V(\phi)\right],
\end{equation}
which are obtained using equation (\ref{16}) under the assumption
that the bulk scalar field is constant with respect to $y$ on the
brane. In other words, it satisfies the following boundary
condition
\begin{equation}\label{26}
\phi_{,y}|_{y=0} = 0.
\end{equation}
Now let us focus attention on the low energies limit
$\rho_m\rightarrow0$. Equation (\ref{23}) thus becomes
\begin{eqnarray}\label{27}
H^2+\frac{k}{a^2} &=& \frac{1}{2r_c^2{\cal
L}'(R)^2}\left[1+\frac{1}{3}r_c^2{\cal L}'(R) \left(R {\cal
L}'(R)-{\cal L}(R)-6H\dot{R}{\cal
L}''(R)\right)\right]\nonumber\\
&+&\frac{\varepsilon}{2r_c^2{\cal L}'(R)^2}\sqrt{1-
\frac{4}{3}\kappa_5^2r_c^2{\cal
L}'(R)^2\rho_{\phi}+\frac{2}{3}r_c^2{\cal L}'(R)\left(R {\cal
L}'(R)-{\cal L}(R)-6H\dot{R}{\cal L}''(R)\right)}.
\end{eqnarray}
Defining $\hat{r}_c=r_c{\cal L}'(R)$ and the curvature energy
density as
\begin{eqnarray}\label{}
\rho_{curv} = \frac{\mu^2}{2{\cal L}'(R)}\left(R {\cal
L}'(R)-{\cal L}(R)-6H\dot{R}{\cal L}''(R)\right),
\end{eqnarray}
the DGP Friedmann equation (\ref{27}) with $k=0$ can be rewritten
as
\begin{equation}\label{phi1}
H^2 = \frac{1}{2\hat{r}_c}+\frac{1}{3\mu^2}\rho_{curv}
+\frac{\varepsilon}{2\hat{r}_c}\sqrt{1-\frac{4}{3}\kappa_5^2\hat{r}_c^2
\rho_{\phi}+\frac{4}{3\mu^2}\hat{r}_c^2\rho_{curv}}.
\end{equation}
In order to investigate the behavior of the solutions in this
modified DGP brane in the presence of the bulk scalar field, we
consider two branches of the solutions under the condition that
$\kappa_5^2\rho_{\phi}\ll\frac{1}{\hat{r}_c^2}$. After obtaining
these solutions, we will check whether this condition is valid or
not. Under such a condition equation (\ref{phi1}) reduces to
\begin{equation}\label{m}
H^2 = \frac{1}{2\hat{r}_c}+\frac{1}{3\mu^2}\rho_{curv}
+\frac{\varepsilon}{2\hat{r}_c}\sqrt{1+\frac{4}{3\mu^2}\hat{r}_c^2\rho_{curv}}.
\end{equation}
This is as far as one could go without specifying the form of ${\cal
L}(R)$. For ease of exposition and clarity, let us focus attention
on theories where the term $R^n$ is present in the brane action and
write
\begin{equation}\label{29}
{\cal L}(R) = R^n.
\end{equation}
For having an accelerated expansion we choose the brane scale
factor as
\begin{equation}\label{30}
a(t) = a_0 e^{\alpha(t-t_0)},
\end{equation}
where $\alpha$ is a positive constant. The solutions with
$\alpha=0$ are not interesting since they provide static
cosmologies with a non-evolving scale factor on the brane. The
Ricci scalar on the brane for a spatially flat FRW geometry is
given by
\begin{equation}\label{Ricci}
R = 12H^2+6\dot{H} = 12\alpha^2.
\end{equation}
Substituting equations (\ref {29}) and (\ref {30}) into equation
(\ref{m}) and using equation (\ref{Ricci}) we obtain
\begin{equation}\label{}
\alpha^2 \hat{r}_c^2 (2-n) = \pm\alpha n.
\end{equation}
where $\hat{r_c}=r_cn12^{n-1}\alpha^{2(n-1)}$. First let us
consider the case $n=1$, namely ${\cal L}(R)=R$. In this case $\rho_{curv}=0$, $\hat{r_c}=r_c$ and we obtain two branches as\\
$\bullet$ $\varepsilon=+1$
\begin{equation}\label{}
H = \alpha = \frac{1}{r_c},
\end{equation}
$\bullet$ $\varepsilon=-1$
\begin{equation}\label{}
H = \alpha =0.
\end{equation}
Therefore, the positive branch is the self-accelerating solution
\cite{Deffayet}. In this case the above condition reduces to
$\kappa_5^2\rho_{\phi}\ll{1}/{r_c^2}$ in agreement with
\cite{India}. In the case $n=2$, we obtain $H=\alpha=0$ in both
branches, that is, a static
universe. For $n>2$ case, the two branches are \\
$\bullet$ $\varepsilon=+1$
\begin{equation}\label{}
H = \alpha =0,
\end{equation}
$\bullet$ $\varepsilon=-1$
\begin{equation}\label{}
H = \alpha =
\left[-\frac{1}{(2-n)r_c12^{n-1}}\right]^{\frac{1}{(2n-1)}}.
\end{equation}
Therefore, under the condition $\kappa_5^2
\rho_{\phi}\ll\left[{1}/{n^{2n-1}(n-2)^{2-2n}r_c12^{n-1}}\right]^{{2}/{(2n-1)}}$
the negative branch has a self-accelerating phase. Since the bulk
scalar field in the absence of brane matter satisfies the usual
momentum conservation law on the brane, we have $\rho_{\phi}\propto
e^{-3\alpha(w_{\phi}+1)t}$. Thus the energy density of matter goes
to zero for late times and reaches a regime where it is small in
comparison with
$\left[{1}/{n^{2n-1}(n-2)^{2-2n}r_c12^{n-1}}\right]^{{2}/{(2n-1)}}$. For the case $n<0$, we have two branches as\\
$\bullet$ $\varepsilon=+1$
\begin{equation}\label{}
H = \alpha =
\left[\frac{1}{(2-n)r_c12^{n-1}}\right]^{\frac{1}{(2n-1)}},
\end{equation}
$\bullet$ $\varepsilon=-1$
\begin{equation}\label{}
H = \alpha =0.
\end{equation}
Therefore, under the condition $\kappa_5^2
\rho_{\phi}\ll\left[{1}/{n^{2n-1}(2-n)^{2-2n}r_c12^{n-1}}\right]^{{2}/{(2n-1)}}$,
the positive branch has a self-accelerating phase. We have
summarized the results in table 1.

In most brane-world models, the $5D$ bulk space-time only includes a
cosmological constant, and the matter fields on the brane are
regarded as responsible for the dynamics of the brane. In this paper
we have shown that in the framework of a modified DGP brane the late
time behavior of the universe does not change even if we ignore the
local matter fields and consider a model of the universe filled with
a bulk scalar field.

An interesting feature of DGP models is the existence of the
ghost-like excitations \cite{g1,g2,g3}. It would therefore be
interesting to mention recent results relevant to the present work.
The study of the spectrum of gravitational perturbations without
matter perturbation about a de Sitter brane shows that for the
positive tension brane the massive spin-2 perturbations contain a
helicity-0 mode that becomes a ghost if the mass is in the range
$0<m^2<2H^2$, while for a negative tension brane the spin-0 mode
becomes a ghost if $Hr_c>1/2$. In the self-accelerating universe
without tension, the mass of the discrete mode of the spin-2
perturbations becomes $2H^2$. This is a special mass in Pauli-Fierz
massive gravity theory because there exists an enhanced symmetry
that eliminates the helicity-0 mode. However, in DGP models, there
is a spin-0 perturbation of the same mass and this breaks the
symmetry, leading to a ghost from the mixing between the spin-0 and
spin-2 perturbations \cite{g2}. It should be mentioned that this is
consistent with the result obtained by the boundary effective action
in \cite{g1} where a scalar mode is found to be a ghost if
$Hr_c>1/2$.

The ghost carries negative energy density and it leads to the
instability of the space-time. The self-accelerating branch of
solutions turns out to be plagued by the ghost instability. In order
to avoid the instability some authors have attempted to construct a
ghost-free model by modifying the self-accelerating branch of the
DGP model \cite{Koyama}. In \cite{Izumi}, the authors have studied
the possibility of avoiding the appearance of ghosts by first
modifying the model via the introduction of a second brane in the
bulk and then stabilizing the brane separation by introducing a bulk
scalar field. It has been shown that it is easy to remove the spin-2
ghost by putting a second brane in the bulk and make the distance
between the two brane small. However it was found that the spin-0
perturbation, the radion, becomes a ghost. By stabilizing the
radion, the perturbations of the scalar field which is necessary to
stabilize the radion becomes a ghost. The most interesting finding
is that it is impossible to remove the spin-2 ghost and spin-0 ghost
simultaneously \cite{Izumi,Charmousis}. However, several interesting
and different ways to avoid the appearance of ghosts is proposed in
\cite{Deffayet}, where it is argued that claims of instability of
the self-accelerating solutions of the DGP model that are based on
linearized calculations are unwarranted. Finally, another
possibility may be to consider altering the theory at the level of
the action itself, for example by adding extra terms in the bulk or
on the brane \cite{g4}. It was also shown that the introduction of
Gauss-Bonnet term in the bulk does not help \cite{new}. In our model
the presence of the ${\cal L}(R)$term, instead of $R$ in the brane
action and the presence of the bulk scalar field provide the new
self-accelerating solutions. Since such a brane effective theory is
definitely a higher derivative theory, there should be a new degree
of freedom which could be a ghost. Probably these new solutions also
suffer from a ghost instability and this could be the subject of a
separate investigation.
\vspace{1.2mm}\noindent\\
\begin{center}
\begin{tabular}{cccccc} \hline\hline${\cal L}(R)=R^n$ &
$\varepsilon=+1$ & $\varepsilon=-1$
\\ \hline\hline
&       &       &       \\
$n<0$  & accelerating universe& static universe \\
$n=1$  & accelerating universe& static universe \\
$n=2$  & static universe      & static universe \\
$n>2$  & static universe      & accelerating universe\\
\\
\hline\hline
\end{tabular}\vspace{2mm}\\
\begin{center}
{\footnotesize Table 1: The late time behavior of the universe for
two branches, positive ($\varepsilon=+1$ ) and negative
$(\varepsilon=-1)$, under the condition
$\kappa_5^2\rho_{\phi}\ll\frac{1}{\hat{r}_c^2}$.}
\end{center}
\end{center}
\vspace{1.2mm}\noindent\\
\section{Conclusions}
In this letter we have derived the effective Einstein field
equations on the brane in the framework of the DGP model where the
action on the brane is an arbitrary function of the Ricci scalar,
${\cal L}(R)$, and the bulk action includes a scalar field in the
matter Lagrangian. We have shown that in a DGP model with curvature
correction, ${\cal L}(R) = R^n$, and a scalar field in the bulk
space, one can obtain an asymptotically static universe and a self
accelerating solution at late times respectively for two different
embeddings of the brane $\varepsilon=-1$ and $\varepsilon=+1$ for
$n=1$ and $n<0$. The role of the two branches is exchanged if we
consider $n>2$. The study of this scenario when the quintom field is
considered as the bulk matter field will be the subject of a future
investigation.

\end{document}